\documentclass[a4paper]{jpconf}
\usepackage{graphicx}
\usepackage{colordvi}
\usepackage{bm}

\usepackage{hyperref}

\usepackage{amssymb}
\usepackage[latin1]{inputenc}
\usepackage{times}
\usepackage{tikz}
\usepackage{amsmath}
\usepackage{verbatim}
\usetikzlibrary{arrows,shapes} 
\usepackage{pifont}
\def\widebar{\overline}
\begin{document}
\title{Semiclassical energy conditions and wormholes}

\author{Prado Mart\'{\i}n-Moruno$^{1,2}$}

\address{$^1$ Centro de Astronomia e Astrof\'{\i}sica da Universidade de Lisboa, Campo Grande, Edif\'{\i}cio C8, 1749-016 Lisboa, Portugal}
\address{$^2$Instituto de Astrof\'{\i}sica e Ci\^{e}ncias do Espa\c{c}o, Universidade de Lisboa, OAL, Tapada da
Ajuda, PT1349-018 Lisboa, Portugal.}

\ead{pmmoruno@fc.ul.pt}

\begin{abstract}
We consider the nonlinear energy conditions 
and their quantum extensions. 
These new energy conditions behave much better than the usual pointwise energy conditions in the presence of semiclassical quantum effects.
Analogous quantum extensions for the linear 
energy conditions are not always satisfied as one enters the quantum realm, but they can be used to constrain the violation of the classical conditions. 
Thus, the existence of wormholes supported by a fluid which violates the null energy condition in a controlled way is of particular interest.
\end{abstract}

\section{Introduction}

As is well known the energy conditions are assumptions made on the material content of the spacetime to extract generic characteristics
of the geometry through the equations of the dynamics (see, for example, references~\cite{H&E,Matt}). 
On one hand, one could assume a {\it reasonable behaviour of matter in a given theory of gravity}. 
A particular example of this approach
is the strong energy condition (SEC) which assumes that gravity is always attractive in General Relativity. That is, requiring the convergence of timelike geodesics and assuming Einstein
equation, one gets that the SEC is the requirement of having a stress-energy tensor which satisfies
\begin{equation}
  {\rm SEC:}\,\,\, T^{a}{}_b\,V_a\,V^b-\frac{1}{2}T\, V^aV_a\geq0,                                                
\end{equation}
for any timelike vector $V^a$.
For a stress-energy tensor that has only one timelike eigenvector (type I in the classification presented in reference~\cite{H&E} ) and, therefore, can be written in an orthonormal basis
as $T^{\hat{a}}{}_{\hat{b}}={\rm diag}\left(-\rho,\,p_1,\,p_2,\,p_3\right)$, the SEC implies
\begin{equation}
 {\rm type\, I:}\,\,\, \rho+p_i\geq0\,\,\&\,\,\rho+\sum p_i\geq0.
\end{equation}
On the other hand, one can also formulate energy conditions by assuming that the {\it material content has some reasonable properties}.
The dominant energy condition (DEC) assumes that the energy density measured by any observer (any timelike vector $V^a$) is non-negative and 
propagates in a causal way. That is, defining the flux vector $F^a=-T^{a}{}_bV^b$ we have
\begin{equation}
 {\rm DEC:}\,\,\, T^a{}_b\,V_a\, V^b\geq0\,\,\&\,\,F^aF_a\leq0\,\,\,\Rightarrow\,\,\,{\rm type\, I:}\,\,\,  \rho\geq |p_i|.
\end{equation}
The weak energy condition (WEC) only requires the energy density measured by any observer to be non-negative, implying
\begin{equation}
 {\rm WEC:}\,\,\,T^a{}_b\,V_a\, V^b\geq0\,\,\,\Rightarrow\,\,\,{\rm type\, I:}\,\,\,\rho\geq 0\,\,\&\,\,\rho+p_i\geq0.
\end{equation}
Finally, the null energy condition (NEC) can be seen as a minimum requirement on the stress-energy tensor.
This is
\begin{equation}
  {\rm NEC:}\,\,\, T^a{}_b\,k_a\, k^b\geq0\,\,\,\Rightarrow\,\,\,{\rm type\, I:}\,\,\,\rho+p_i\geq0,
\end{equation}
where $k^a$ is any null vector. As it is implied by both the WEC (which is a consequence of the DEC) and the SEC, a hypothetical material violating this condition would not satisfy any of the aforementioned
conditions.

These energy conditions have been proven to be of great utility to extract conclusions about generic geometries \cite{H&E,Matt}. Moreover, 
they have also been used to argue that some solutions of the Einstein equations
with potential pathological characteristics should be discarded, since they are supported by a matter content violating some of these conditions.
That is the case for wormholes \cite{Matt}, which are shortcuts between different regions of one universe or connections between different universes.
For traversable wormholes the metric can be given in the Morris-Thorne form \cite{Morris:1988cz}
\begin{equation}
 ds^2=-e ^{2\Phi(r)} \,dt^2+\frac{dr^2}{1- b(r)/r}+r^2 \,(d\theta^2+\sin ^2{\theta} \, d\phi ^2)
\end{equation}
where $r_0\leq r\leq\infty$,
$r_0$ is the radius of the throat, $r_0=b(r_0)\leq b(r)$, $\Phi(r)<\infty$ for any value of $r$ to avoid horizons, and 
$b'(r)<b(r)/r$ close to the throat for it to ``flare-out''.
Considering the Einstein equations it can be seen that one would need the existence of exotic matter close to the throat to keep it open, 
which exoticity resides precisely on the fact that it violates the NEC since it has
\begin{equation}
 \rho+p_r<0.
\end{equation}
Thus, if the energy conditions were universal, wormholes would not exist in a universe described by general relativity. (This is not necessarily the case when considering alternative
theories of gravity \cite{mwhs}.)

Nevertheless, some doubts have been raised about this universality.
It is clear that the SEC has to be significantly violated in cosmological scenarios, both during inflation and right now. 
Moreover, the SEC is also violated by semiclassical quantum effects that have been measured in the laboratory.  
Therefore, this condition is almost
abolished.
On the other hand, the other pointwise energy conditions, which are independent of the theory of gravity, are also known to be violated, for example, by the renormalized stress-energy tensor in the Casimir vacuum
(bounded violations) and in the Schwarzschild spacetime (unbounded violations for the Boulware and Unruh vacua \cite{unboundedHH,unboundedB,unboundedU}).

%%%%%%%%%%%%%%%%%%%%%%%%%%%%%%%%%%%%%%%%%%%%%%%%%%%%%%%%%%%%%%%%%%%%%%%%%%%%%%%%%%%%%%
%%%%%%%%%%%%%%%%%%%%%%%%%%%%%%%%%%%%%%%%%%%%%%%%%%%%%%%%%%%%%%%%%%%%%%%%%%%%%%%%%%%%%%
\section{New pointwise energy conditions}
Given the situation presented in the previous section, the main approach that has been developed in the literature consists in averaging 
those energy conditions along timelike or null curves,
or in using some quantum inequalities to restrict the magnitude and duration of negative energies \cite{QI}. 
However, we will follow a different approach, considering other pointwise energy conditions.

%%%%%%%%%%%%%%%%%%%%%%%%%%%%%%%%%%%%%%%%%%%%%%%%%%%%%%%%%%%%%%%%%%%%%%%%%%%%%%%%%%%%%%
%%%%%%%%%%%%%%%%%%%%%%%%%%%%%%%%%%%%%%%%%%%%%%%%%%%%%%%%%%%%%%%%%%%%%%%%%%%%%%%%%%%%%%
\subsection{Nonlinear energy conditions}

The first condition we present was first considered in reference~\cite{FEC} when obtaining entropy bounds for uncollapsed systems, and was latter studied in deeper detail in reference~\cite{QFEC}. This is the
flux energy condition (FEC) which assumes that the energy density measured by any observer propagates in a causal way. That is,  it states \cite{FEC,QFEC}
\begin{equation}
{\rm FEC:}\,\,\, F^aF_a\leq0 \,\,\,\Rightarrow\,\,\,{\rm type\, I:}\,\,\, \rho^2\geq p_i^2.
\end{equation}
Therefore, by their very definition, the WEC plus the FEC is equal to the DEC. 
As some encouraging results were obtained for this condition in relation to the Boulware vacuum of a Schwarzschild black hole (as we will summarize
in the next section), in reference~\cite{nonlinear} other nonlinear energy conditions were formulated. These are the determinant energy condition (DETEC) and
the trace-of-square energy condition (TOSEC), which are given by \cite{nonlinear}
\begin{equation}
 {\rm DETEC:}\,\,\, \det \left( T^{ab}\right) \geq 0\,\,\,\Rightarrow\,\,\,{\rm type\, I:}\,\,\, \rho\cdot\prod p_i\geq0,
\end{equation}
and
\begin{equation}
  {\rm TOSEC:}\,\,\, T^{ab}\, T_{ab} \geq 0\,\,\,\Rightarrow\,\,\,{\rm type\, I:}\,\,\, \rho^2+\sum p_i^2\geq0,
\end{equation}
respectively. Therefore, the TOSEC is always satisfied if the stress-energy tensor is type I. Indeed, it can only be violated for type IV stress-energy tensors \cite{nonlinear}. 
Although the physical interpretation of these conditions is not so clear as in the case of the FEC, we present them for completeness.

%%%%%%%%%%%%%%%%%%%%%%%%%%%%%%%%%%%%%%%%%%%%%%%%%%%%%%%%%%%%%%%%%%%%%%%%%%%%%%%%%%%%%%
%%%%%%%%%%%%%%%%%%%%%%%%%%%%%%%%%%%%%%%%%%%%%%%%%%%%%%%%%%%%%%%%%%%%%%%%%%%%%%%%%%%%%%
\subsection{Semiclassical energy conditions}
Following a different point of view, one can note that the violations of the energy conditions which have been measured in the laboratory are not arbitrarily large. 
Therefore, one could formulate new pointwise energy conditions
based on considering bounded violations of the energy conditions for the renormalized stress-energy tensor in semiclassical scenarios, as it has been done in references \cite{QFEC,nonlinear}.

In the first place, we consider the quantum flux energy condition (QFEC) which states that the energy density measured by any observer either 
propagates in a causal way or does not propagate ``too much''. That is, this condition assumes that 
either the flux vector is non-spacelike or at least its norm is bounded by above
by a bound that depends on the characteristics of the system: number of fields ($N$), system 4-velocity ($U^a$), and a characteristic distance ($L$). A possible formulation is \cite{QFEC}
\begin{equation}
{\rm QFEC:}\,\,\, F^aF_a \leq \zeta \;(\hbar N/L^4)^2\; (U_a\,V^a)^2\,\,\,\Rightarrow\,\,\,{\rm type\, I:}\,\,\, \rho^2-p_i^2\geq-\zeta \;(\hbar N/L^4)^2,
\end{equation}
where $\zeta$ is a parameter of order unity.
In the second place, similar quantum generalizations of the other nonlinear conditions have been formulated \cite{nonlinear}.
Moreover, one can also consider similar quantum extensions for the linear energy conditions. For example, the quantum weak energy 
condition (QWEC)
assumes that the energy density measured by any observer should not be ``excessively negative'', that is \cite{nonlinear}
\begin{equation}
{\rm QWEC:}\,\,\,T^{a}{}_b\, V_a V^b \geq -\zeta\hbar N/L^4\; (U_a\,V^a)^2\,\,\,\Rightarrow\,\,\,{\rm type\, I:}\,\,\, \rho+p_i\geq-\zeta \;\hbar N/L^4.
\end{equation}

%%%%%%%%%%%%%%%%%%%%%%%%%%%%%%%%%%%%%%%%%%%%%%%%%%%%%%%%%%%%%%%%%%%%%%%%%%%%%%%%%%%%%%
%%%%%%%%%%%%%%%%%%%%%%%%%%%%%%%%%%%%%%%%%%%%%%%%%%%%%%%%%%%%%%%%%%%%%%%%%%%%%%%%%%%%%%
\section{Quantum vacuum states and energy conditions}

Now, let us summarize the behaviour of the renormalized stress-energy tensor of some vacuum states with respect to the energy conditions (see reference~\cite{nonlinear} for further details). As it is included in table \ref{t1},
the quantum extensions of the linear and nonlinear energy conditions are satisfied in the Casimir vacuum taking the distance between the parallel plates as the characteristic distance of the system. 
Moreover, the DETEC and TOSEC are also satisfied.

\begin{table}[h]
\caption{\label{t1}Casimir spacetime. \color{red}{v} \color{black} and \color{green}{s} \color{black} 
denote that the condition is violated and
satisfied, respectively.}
\begin{center}
\begin{tabular}{*{6}{l}}
\br
WEC&QWEC&FEC&QFEC&DETEC&TOSEC\\
\mr
\color{red}{v} & \color{green}{s} & \color{red}{v} & \color{green}{s} & \color{green}{s} & \color{green}{s} \\
\br
\end{tabular}
\end{center}
\end{table}

On the other hand, we consider the renormalized stress-energy tensor of a massless conformally coupled scalar field theory in a Schwarzschild geometry obtained
taking Page's analytic approximation \cite{Page} for the Hartle--Hawking vacuum,
combining that approximation with the results of Brown and Ottewill \cite{Brown,unboundedB} for the Bolware vacuum, and Visser's semi-analytical approximation for the Unruh vacuum \cite{unboundedU}. 
The results are summarized in table \ref{t2}. As the Hartle--Hawking vacuum entails only finite quantities, one could expect a better behaviour for the quantum extensions of the
energy conditions, and this is the case taking the characteristic distance to be the black hole radius (which is a natural assumption). 
It should be noted that the quantum nonlinear energy conditions are satisfied in all the cases.
One can wonder why the TOSEC is violated in the Unruh vacuum since it is fulfilled by any type I stress-energy tensor. The reason is because this renormalized 
stress-energy tensor is of type IV in the asymptotic region \cite{nonlinear}.

\begin{table}[h]
\caption{\label{t2}Schwarzschild geometry: Massless conformally coupled scalar.
 \color{red}{v} \color{black} and \color{green}{s} \color{black} 
denote that the condition is violated and
satisfied, respectively.}
\begin{center}
\begin{tabular}{*{9}{l}}
\br
 &WEC&QWEC&FEC&QFEC&DETEC&QDETEC&TOSEC&QTOSEC\\
\mr
Hartle--Hawking& \color{red}{v} & \color{green}{s} & \color{red}{v} &\color{green}{s} & \color{red}{v} & \color{green}{s} &\color{green}{s} &\color{green}{s}\\
Boulware& \color{red}{v} & \color{red}{v} & \color{green}{s} & \color{green}{s} & \color{red}{v} & \color{green}{s} &\color{green}{s} &\color{green}{s}\\
Unruh&\color{red}{v} & \color{red}{v} & \color{red}{v} & \color{green}{s} & \color{red}{v} & \color{green}{s} &\color{red}{v} &\color{green}{s}\\
\br
\end{tabular}
\end{center}
\end{table}

%%%%%%%%%%%%%%%%%%%%%%%%%%%%%%%%%%%%%%%%%%%%%%%%%%%%%%%%%%%%%%%%%%%%%%%%%%%%%%%%%%%%%%
%%%%%%%%%%%%%%%%%%%%%%%%%%%%%%%%%%%%%%%%%%%%%%%%%%%%%%%%%%%%%%%%%%%%%%%%%%%%%%%%%%%%%%
\section{Wormholes satisfying the QWEC}
As we have already pointed out in order to have a wormhole geometry one would need some amount of exotic matter.
Due to the violation of the NEC by this material, it has been considered how to
minimize the quantity of exotic matter needed to support a wormhole by considering thin-shells to construct the tunnel itself 
(see \cite{Garcia:2011aa} and references therein)
or to surround the mouth of a traversable wormhole matching 
it to a spacetime where no exotic matter is present \cite{Lobo:2005zu}. 
It must be noted that although the NEC can be violated by semiclassical quantum effects, unbounded violations are only associated to some black hole vacua (where the stress-energy tensor itself diverges).
Therefore, a different (and compatible) approach to reduce the exoticity of these geometries has been 
followed in reference~\cite{whs}, that is
to minimize the violation of the NEC considering that only small bounded violations of this
energy condition are possible. Thus, one would minimize not only the quantity of exotic stuff to be used but also its exoticity.

%%%%%%%%%%%%%%%%%%%%%%%%%%%%%%%%%%%%%%%%%%%%%%%%%%%%%%%%%%%%%%%%%%%%%%%%%%%%%%%%%%%%%%
%%%%%%%%%%%%%%%%%%%%%%%%%%%%%%%%%%%%%%%%%%%%%%%%%%%%%%%%%%%%%%%%%%%%%%%%%%%%%%%%%%%%%%
\subsection{Equation of state minimally violating the WEC}
The equation of state that we will take as a starting point has been first considered in a cosmological setting when investigating how a small constant deviation from a cosmological constant would
change the future of the universe to what has been called {\it the little sibling of the big rip} \cite{little}. That equation of state is 
\begin{equation}
 p_{\rm DE}+\rho _{\rm DE}\ =\ -\frac{A}{3}.
\end{equation}
As $\rho_{\rm DE}\geq0$, the QWEC would be satisfied for $A \ll \hbar H_0^4$ \cite{little}.

For wormhole geometries an anisotropic version of this equation of state has been considered in reference~\cite{whs}. This is
\begin{equation}
 p_r(r)+\rho _{r}(r)\ =\ -\frac{A}{8\pi},
\end{equation}
with  $\rho(r)\geq0$ and  $A \ll 1/r_0^2$. Although there cannot be asymptotically flat solutions with this equation of state 
(since it has to be also satisfied in the asymptotic limit), 
this kind of solutions has been constructed considering thin-shells. 
It has been proven that the resulting construction is stable and satisfies the QWEC, minimizing also the quantity of material violating the WEC \cite{whs}.
On the other hand, one can also consider an inhomogeneous version of this equation of state such as 
$\rho+p_r\rightarrow -\frac{A}{8\pi}$ when $r\rightarrow r_0$ and
$\rho+p_r\rightarrow 0$ when $r\rightarrow\infty$,
which is not only compatible with asymptotic flatness (at least in principle), but minimizes even more the violation
of the WEC without the introduction of thin-shells. In reference~\cite{whs} the following equation has been considered:
\begin{equation}\label{estado}
 \rho(r)+p_r(r)=-\frac{A}{8\pi}\left(\frac{r_0}{r}\right)^\alpha
\end{equation}
where $\rho(r)\geq0$ and  $A \ll 1/r_0^2$.

%%%%%%%%%%%%%%%%%%%%%%%%%%%%%%%%%%%%%%%%%%%%%%%%%%%%%%%%%%%%%%%%%%%%%%%%%%%%%%%%%%%%%%
%%%%%%%%%%%%%%%%%%%%%%%%%%%%%%%%%%%%%%%%%%%%%%%%%%%%%%%%%%%%%%%%%%%%%%%%%%%%%%%%%%%%%%
\subsection{Particular solution}
Considering equation of state (\ref{estado}) with $\alpha>2$ and assuming a constant shift function $\Phi(r)=\Phi_0$, one can obtain the shape function $b(r)$ and the transverse pressure through the Einstein
equations. The metric describing this geometry is \cite{whs}
\begin{equation}
 {\rm d}s^2=-e^{2\Phi_0}{\rm d}t^2+\frac{\alpha-2}{A\,r_0^2}\frac{{\rm d}r^2}{1-\left(r_0/r\right)^{\alpha-2}}+r^2 {\rm d}\Omega_{(2)}^2.
\end{equation}
Thus, close to $r_0\lesssim r$ we have $b'(r)<b(r)/r$, that is a wormhole throat. In the asymptotic limit the geometry tends to
\begin{equation}
 d\widebar s^2=-d\widebar{t}^2+d\widebar{r}^2+\frac{A\,r_0^2}{\alpha-2}\widebar{r}^2 d\Omega_{(2)}^2,
\end{equation}
where $\widebar r^2=\frac{(\alpha-2)r^2}{A\,r_0^2}$ and $\widebar t=e^{\Phi_0}t$. 
The area of the sphere of radius $\widebar r$ is $4\pi(1-\Delta)\widebar r^2$, with
$\Delta=1-\frac{A\,r_0^2}{\alpha-2}$. Thus, this geometry describes a space with a deficit of solid angle \cite{monopolo} in 
the asymptotic limit if $\alpha>3$ and if $2<\alpha<3\,\,\&\,\,Ar_0^2<\alpha-2$.
This solution may, therefore, be interpreted as a wormhole carrying a global monopole \cite{whs}.

The energy density and pressures of the material content can be obtained using the Einstein equations. These are \cite{whs}
\begin{eqnarray}
 \rho&=&\frac{1}{8\pi(\alpha-2)r^2}\left\{\alpha-2-Ar_0^2\left[1+(\alpha-3)
\left(\frac{r_0}{r}\right)^{\alpha-2}\right]\right\}\\
p_r&=&-\frac{1}{8\pi(\alpha-2)r^2}\left\{\alpha-2-Ar_0^2
\left[1-\left(\frac{r_0}{r}\right)^{\alpha-2}\right]\right\}\\
p_t&=&\frac{A}{16\pi}\left(\frac{r_0}{r}\right)^{\alpha}>0.
\end{eqnarray}
Therefore, $p_t(r)\geq0$. As $\rho(r)\geq0$ for $\alpha>3$ and $2<\alpha<3\,\,\&\,\,Ar_0^2<\alpha-2$, we would have $\rho(r)+p_t(r)>0$ and, considering also equation (\ref{estado}), the QWEC is satisfied through
the whole space.

%%%%%%%%%%%%%%%%%%%%%%%%%%%%%%%%%%%%%%%%%%%%%%%%%%%%%%%%%%%%%%%%%%%%%%%%%%%%%%%%%%%%%%
%%%%%%%%%%%%%%%%%%%%%%%%%%%%%%%%%%%%%%%%%%%%%%%%%%%%%%%%%%%%%%%%%%%%%%%%%%%%%%%%%%%%%%
\section{Discussion}

We have reviewed how the quantum generalizations of the energy conditions behave better than their classical counterparts. Moreover, the nonlinear energy conditions are satisfied in more situations than
the usual energy conditions.
The QFEC is satisfied for classical situations and quantum vacuum states.

Regarding the applicability of these new energy conditions, the FEC has been used to obtain some interesting entropy bounds for 
uncollapsed systems \cite{FEC}. As the QFEC is satisfied when Hawking evaporation is present, one should not expect this condition to be compatible with a derivation of the second law of black hole physics. Although its utility 
for other theoretical problems is completely unexplored for the moment, we can already argue that some steps in its application to 
generic spacetimes would be particularly tricky 
due to both the nonlinear character of this condition and its non-vanishing
r.~h.~s. It has already been pointed out \cite{nonlinear} that one could think that two quantum field theories satisfying the QFEC may interfere in a destructive way under some circumstances 
due to the nonlinear character of this condition and produce
a final situation where it is violated, although an explicit example of such behaviour have not been presented yet.

Inspired by the results of the Casimir vacuum, one may think that the quantum linear energy conditions could be enough for describing semiclassical phenomena, giving a preferred status to the Hawking--Hartle vacuum
for the black hole space. In this case, it would not only be interesting to study the possible existence of wormhole geometries satisfying the QWEC (as it has been done in reference~\cite{whs}), 
but also to explore the utility of this condition to deduce characteristics of a generic spacetime where it is satisfied. 
Moreover, in the case that this utility may be proven, it would be of particular interest to check
if the mentioned wormhole geometries could avoid pathologies (as to be used as hypothetical time-machines) as long as they continue satisfying the QWEC.

\ack
I want to thank Matt Visser for enlightening discussions and co-authorship of references~\cite{QFEC,nonlinear}, and
Mariam Bouhmadi-L\'opez and Francisco S.~N.~Lobo for collaboration on reference~\cite{whs}, papers in which this talk was mainly based on.
This work was supported by FCT (Portugal) through the projects PTDC/FIS/111032/2009 and EXPL/FIS-AST/1608/2013.

\end{document}